\RequirePackage{fix-cm}
\documentclass[smallextended,final]{svjour3mod}       
\smartqed  
\usepackage{cite}
\usepackage{graphicx}
\usepackage{bbding}
\usepackage{multirow}
\usepackage{amsmath,amssymb}
\usepackage{hyperref}
\usepackage{xcolor,soul}
\sethlcolor{yellow}

\colorlet{hyper_colour}{red!70!black}
\hypersetup{
    colorlinks=true,
    linkcolor=hyper_colour,
    urlcolor=hyper_colour,
    filecolor=hyper_colour,
    citecolor=hyper_colour,
}
\usepackage{doi}

\usepackage{siunitx}
\DeclareSIUnit{\gauss}{G}

\usepackage{makecell}
 \journalname{Few-Body Syst}

\newcommand{\Rb}[1]{${}^{#1}$Rb}
\newcommand{\K}[1]{${}^{#1}$K}

\newcommand{\ket}[1]{\ensuremath{\left|#1\right>}}

\newcommand{\etal}{{\it et al.\/}}
\newcommand{\ie}{{\it i.e.\/}}

\newcommand{\ii}{\mathrm{i}}
\newcommand{\ee}{\mathrm{e}}
\newcommand{\dd}{\ensuremath{\mathrm{d}}}

\newcommand{\eref}[1]{Eq.~(\ref{#1})}

\newcommand{\fref}[1]{Fig.~\ref{#1}}
\newcommand{\figref}[1]{Fig.~\ref{#1}}

\newcommand{\Figref}[1]{Figure~\ref{#1}}

\newcommand{\sref}[1]{section~\ref{#1}}

\newcommand{\tref}[1]{table~\ref{#1}}

\newcommand\todo[1]{{#1}}

\usepackage{mathtools}
\DeclareMathOperator{\real}{Re}
\DeclareMathOperator{\imag}{Im}

\begin{document}

\title{Observing S-Matrix Pole Flow in Resonance Interplay
}
\subtitle{Cold Collisions of Ultracold Atoms in a Miniature Laser-based Accelerator}


\author{Matthew Chilcott  \and Samyajit Gayen \and James Croft \and Ryan Thomas \and Niels Kj{\ae}rgaard 
}


\institute{Matthew Chilcott (\Envelope) \and Samyajit Gayen \and Niels Kj{\ae}rgaard (\Envelope) \at
              Department of Physics, QSO—Quantum Science Otago, and Dodd-Walls Centre for Photonic and Quantum Technologies,
University of Otago, Dunedin 9016, New Zealand\\
              \email{matthew.chilcott@otago.ac.nz}         \email{niels.kjaergaard@otago.ac.nz} 
           \and
           James Croft \at
              Joint Quantum Centre (JQC) Durham-Newcastle, Department of Chemistry, Durham University, Durham DH1 3LE, United Kingdom
              \and
           Ryan Thomas \at
               Department of Quantum Science and Technology, Australian National University, Canberra, Australia
}

\date{Dated: \today}

\maketitle
\begin{abstract}
We provide an overview of experiments exploring resonances in the collision of ultracold clouds of atoms. Using a laser-based accelerator that capitalizes on the energy resolution provided by the ultracold atomic setting, we unveil resonance phenomena such as Feshbach and shape resonances in their quintessential form by literally photographing the halo of outgoing scattered atoms. We exploit the tunability of magnetic Feshbach resonances to instigate an interplay between scattering resonances. By experimentally recording the scattering in a parameter space spanned by collision energy and magnetic field, we capture the imprint of the $S$-matrix pole flow in the complex energy plane. After revisiting experiments that place a Feshbach resonance in the proximity of a shape resonance and an anti-bound state, respectively, we discuss the possibility of using $S$-matrix pole interplay between two Feshbach resonances to create a bound-state-in-the-continuum.
\end{abstract}

\section{Introduction}
\label{intro}
Resonance phenomena are ubiquitous in physics, appearing in all manner of mechanical, electrical, acoustic, optical and quantum mechanical systems. In theories of quantum scattering, one way to parameterise resonant scattering between particles is by means of poles of the system's analytically-continued $S$ matrix~\cite{ParticleReview}. While these poles reside at physically unreachable complex energies and have at times been considered ``mathematical oddities''~\cite{McVoy1967}, their effect can nevertheless be seen on the real, positive energy axis where experiments are conducted~\cite{McVoy1967,Ceci2017,ParticleReview,AhmedBraun2021}. Since the seminal work by Nussenzveig~\cite{Nussenzveig1959} that studied the flow of $S$-matrix poles while modifying the depth of a square well potential, multiple authors have greatly expanded the theoretical understanding of $S$-matrix poles and their trajectories~\cite{Potvliege1988,Dabrowski1997,Belchev2011,Racz2011,Meeten2019,Simon2019,Ershov2021}. Observations of $S$-matrix pole flow in scattering experiments have, however, been somewhat wanting as the interaction potentials describing collisions between material particles are typically not tunable. 

\todo{This contribution is based on an invited plenary talk presented at the 25th European Conference on Few-Body Problems in Physics. Here one us (N.K.) gave an overview of experiments with a minature laser-based collider, which has been in operation in New Zealand since 2012~\cite{Rakonjac2012}. In this vein, we shall in the below consider scattering experiments using $\rm ^{87}Rb$ and $\rm ^{40}K$, which are bosonic and fermionic workhorses, respectively, of ultracold atomic physics as these alkali species are readily laser cooled~\cite{Inguscio2013}. When augmented with evaporative cooling~\cite{Deb2014} atomic samples with temperatures less than a microkelvin may be obtained, and these can serve as targets and projectiles in a collision experiment}. Moreover, $\rm ^{87}Rb$ and $\rm ^{40}K$ possess mag\-netically-tunable Feshbach resonances in their inter- and intra-atomic interactions. Utilising the magnetic tuning, the $S$-matrix resonance pole associated with a Feshbach resonance can be brought in proximity to other poles of the $S$ matrix. These other poles may arise in the entrance channel due to bound or anti-bound states, shape resonances or additional Feshbach resonances. Numerical calculations of $S$-matrix elements are feasible thanks to accurate models of the interaction potentials for Rb-Rb~\cite{Strauss2010} and Rb-K~\cite{Pashov2007} and to the reduced number of quantum states involved in the collision at ultra-low energies.
We interpret our experimental observations in terms of pole flow of the analytically-continued $S$ matrix. In particular, we demonstrate that the numerically-ascertained pole positions and their trajectories have a conspicuous impact on our atomic scattering experiments. \todo{As an outlook towards future experiments, we consider the case of two coupled Feshbach resonances which during their pole-flow establish a bound-state-in-the-continuum.}

\section{Experimental setup}

\begin{figure}
  \includegraphics[width=\textwidth]{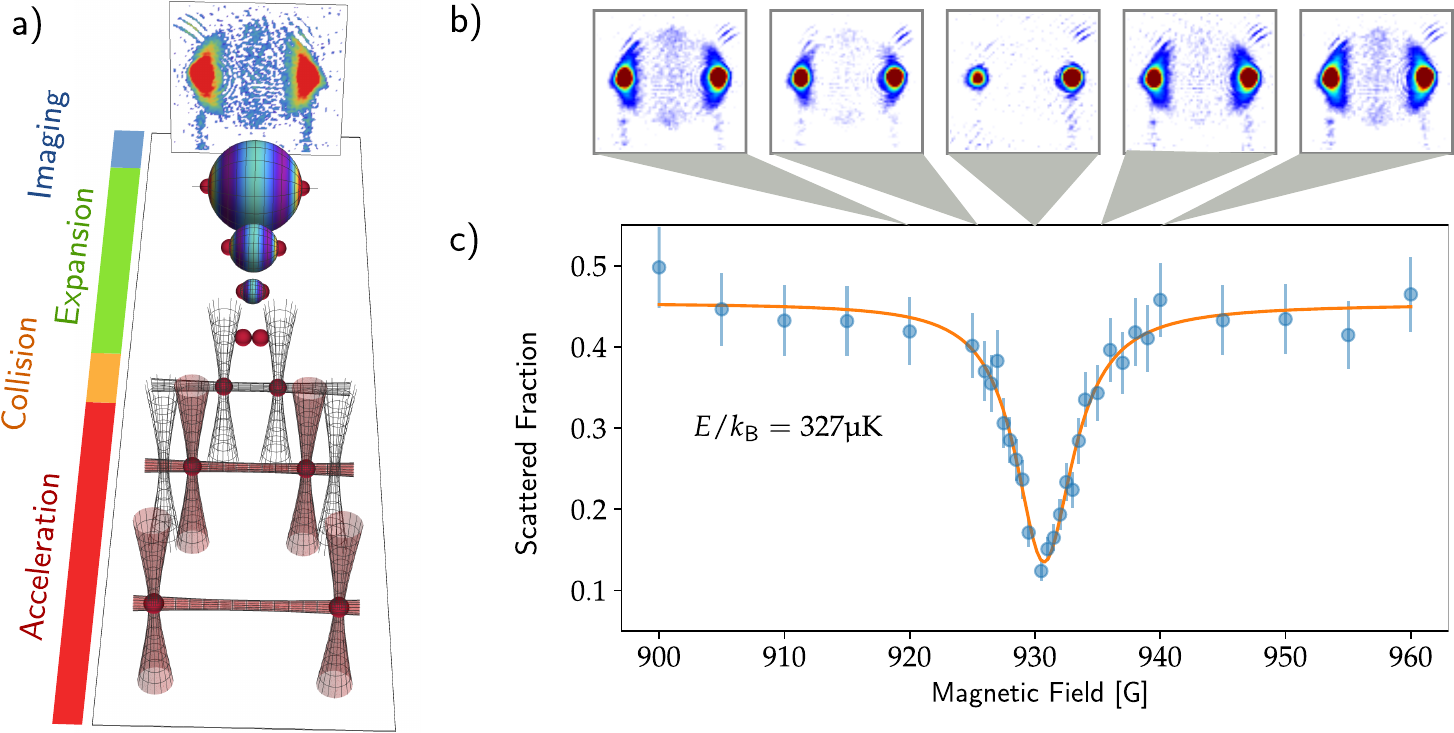}
    \caption{Cold collisions of ultracold atoms. a) The collider sequence: acceleration in optical tweezers, collision in the absence of trapping, ballistic expansion of the scattering halo, and, finally, imaging of said halo and unscattered atoms. b) A series of images of \Rb{87} atoms collected for a range of magnetic fields at a fixed energy, demonstrating the variation of scattering strength imparted by a Feshbach resonance. c) The fraction of incoming atoms scattered, for the fixed energy of \SI{327}{\micro\kelvin} over a range of fields, as extracted from images including those shown in b). The data presented in c) were previously published in Ref.~\cite{Chilcott2021}.}
    \label{fig:overview}
\end{figure}
\Figref{fig:overview}a summarises the sequential operation of our miniature atomic collider. Briefly, two ultracold clouds of atoms (shown as red spheres) are held in the crossings of two vertical laser beams with a horizontal laser beam. The atoms are confined at points of high laser intensity---the beam crossings---by the optical dipole force~\cite{Grimm2000}. They are loaded into these optical tweezers from a magnetic trap where evaporative cooling down to a temperature of \SI{\sim 200}{\nano\kelvin} can be performed. The optical tweezer system manipulates the two clouds of atoms into colliding by steering the vertical beams as shown in the acceleration phase of \figref{fig:overview}a. The two potential wells confining the atoms are the result of rapidly toggling the position of a single beam through the frequency drive of an acousto-optic deflector~\cite{Chisholm2018}. Once the clouds are accelerated into their collision course, the confining laser beams are turned off so the atoms can collide in free space without external influence other than a chosen uniform magnetic field. This miniature collider setup accelerates each cloud over a maximum distance of \SI{3}{\milli\metre}, to explore a domain of collision energies around hundreds of nano-eV. These energies are in the ``cold'' domain with $E/k_\text{B}$ reaching up to about a millikelvin while the clouds of atoms themselves are ``ultra-cold'' with temperatures of $\lesssim$\SI{1}{\micro\kelvin}~\cite{Weiner1999}.

After the collision, the clouds and the scattering halo are allowed to expand for a few milliseconds, and the uniform magnetic field is turned off. A laser pulse then projects a shadow image of the atoms onto a CCD camera (see `imaging' in \figref{fig:overview}a). \Figref{fig:overview}b shows examples of such laser absorption images of scattering halos, acquired at a fixed energy $E/k_B=\SI{327}{\micro\kelvin}$ for range of externally applied $B$-fields. From the analysis of such images~\cite{Kjaergaard2004}, the partial-wave components of the scattering and the scattered fraction can be extracted. The latter is plotted \figref{fig:overview}c and displays a dramatic extinction at $B\sim 930$~G as a result of the destructive interference between a $d$-wave shape resonance and a $d$-wave Feshbach resonance~\cite{Chilcott2022}. 
\todo{
\section{\texorpdfstring{$S$}{S}-matrix poles and their interplay}

\begin{figure}
\includegraphics[width=\linewidth]{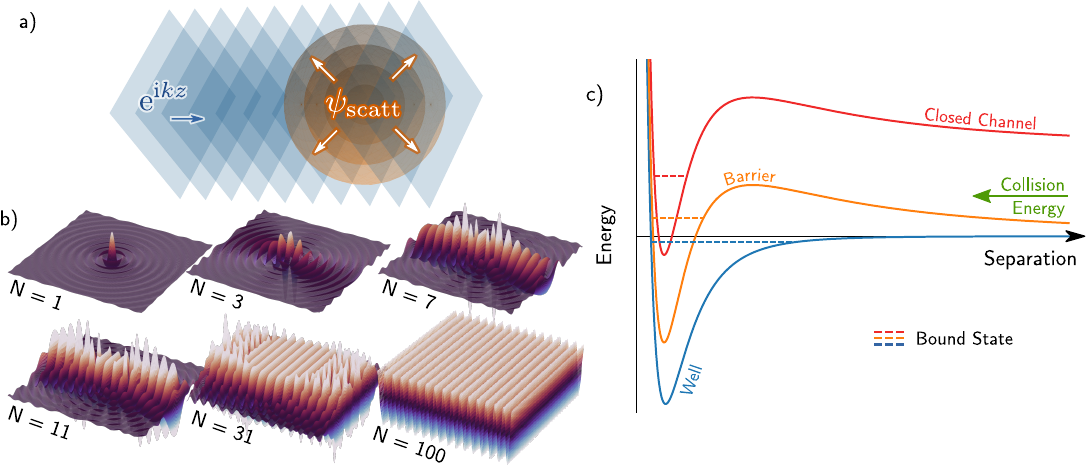}
  \caption{\todo{a) Graphical representation of \eref{eq:plsphe} describing scattering as a sum of a plane wave traveling in the $z$-direction and spherical wave propagating outwards from the origin. b) Partial sums of the Rayleigh expansion, illustrating how a plane wave is formed as a superposition of spherical waves in the limit $N \to \infty$. c) (Quasi-)bound states in scattering potentials. The well of the interaction potential in the open entrance channel (blue) may host an (anti-)bound state below threshold, producing a sub-threshold resonance. The addition of angular momentum to the open channel introduces a centrifugal barrier (orange) which can host the quasi-bound state of a shape resonance. A closed channel (red) can support a bound-state, coupling to which produces a Feshbach resonance.}}
  \label{fig:scattering}
\end{figure}
To set the scene, we first consider the elastic collisions between two particles. Standard textbook treatments of time-independent quantum scattering transform this into the equivalent problem of a single incoming particle of reduced mass $\mu$ and energy $E = \hbar^2 k^2 / 2 \mu$ scattering off a potential localised at the origin~\cite{Friedrich2013,Taylor2006}. The incoming particle is represented by a plane wave, $\ee^{\ii k z}$ of momentum $\hbar k$ along the $z$ axis.
At long-range far away from the origin, the scatted wavefunction takes the form of a spherical wave $\psi_\text{scatt}$ propagating radially outward. The outgoing wave is angularly modulated by the scattering amplitude $f(\theta, \phi)$. It is the scattering potential that determines the scattering amplitude in a given direction and if the potential is radially symmetric, $f$ will be independent of the azimuthal angle $\phi$. The total stationary wavefunction then has the asymptotic form, 
\begin{equation}\label{eq:plsphe}
\psi \overset{r\to\infty}{\sim} {\ee^{\ii k z}} + \underbrace{f(k,\theta)\frac{\ee^{\ii k r}}{r}}_{\psi_\text{scatt}}, 
\end{equation}
which is depicted in \figref{fig:scattering}a.
Inspired by the spherical nature of the scattered wavefunction, the plane wave is expressed as linear combination of spherical waves via the Rayleigh expansion
\begin{equation}\label{eq:rayleigh}
    \ee^{\ii k z} = \sum_{\ell = 0}^{\infty} (2\ell + 1)\ii^\ell j_\ell(kr)P_\ell (\cos\theta),
\end{equation}
where $j_\ell$ are spherical Bessel functions of the first kind and $P_\ell$ are Legendre polynomials. The construction of the plane wave from this series is shown in \figref{fig:scattering}b. 

Without a scattering interaction, the total wavefunction of \eref{eq:plsphe} is simply a plane wave, with each $\ell^\text{th}$ component of the expansion \eref{eq:rayleigh} giving the long-range behaviour
\begin{equation}
     \psi_\ell^\text{unscatt}\overset{r\to\infty}{\sim} \frac{ \sin\left(kr - \ell\pi/2\right)}{r}P_\ell (\cos\theta).
\end{equation}
Sharing the azimuthal symmetry of the plane wave and the potential, the total wavefunction including a scattering interaction can be expanded over the same basis of partial waves,
\begin{equation}\label{eqn:radial_expansion}
  \psi =  \sum_{\ell = 0}^{\infty} \frac{u_\ell (r)}{r}P_\ell(\cos \theta).
\end{equation}
Therefore if the two particles do not interact, the radial partial wavefunction $u_\ell(r)$ of two non-interacting particles takes the form
\begin{equation}
  u_{\ell}^\text{unscatt} \overset{r\to\infty}{\sim} \sin\left(kr - \ell\pi/2\right) = \frac{\ii}{2}\left[\ee^{-\ii({kr - \ell\pi/2)}} - \ee^{\ii{(kr - \ell\pi/2)}}\right],
\end{equation}
which is a sum of a radial wave propagating towards the origin, $\propto \ee^{-\ii kr}$, and one propagating outwards, $\propto \ee^{+\ii kr}$. Physically, the incoming wave is fixed, and thus we would expect the scattering to affect only the outgoing wave
\begin{equation}\label{eqn:wavefunc_longrange_boundary}
  u_{\ell} \overset{r\to\infty}{\sim} \frac{\ii}{2}\left[\ee^{-\ii(kr - \ell\pi/2)} - S_\ell \ee^{+\ii(kr - \ell\pi/2)}\right].
\end{equation}
This defines the $\ell^\text{th}$ component of the $S$ matrix. Because of conservation of particle current, the only possible change with respect to the long-range unscattered wave component is a phase-shift $\delta_\ell$, the so-called scattering phase:
\begin{equation}
  u_\ell \overset{r\to\infty}{\sim} \sin(kr - \ell\pi/2 + \delta_\ell),
\end{equation}
which in turn provides an expression for the $S$ matrix
\begin{equation}\label{eq:sell}
   S_\ell = \ee^{2\ii \delta_\ell}.
\end{equation}
Note that the conservation of particle current implies that $S_\ell$ has unit modulus.

In addition to the long-range behaviour of \eref{eqn:wavefunc_longrange_boundary}, a physical wavefunction must be regular at the origin. By inspection of \eref{eqn:radial_expansion} this requires $u_\ell \to 0$ as $r \to 0$. Between these two boundary conditions, the radial wavefunction is a solution to the radial Schr\"odinger equation,
\begin{equation}\label{eqn:radial_schrodinger}
  \left( \frac{\hbar^2}{2 \mu}\frac{\dd^2}{\dd r^2} - \frac{\ell(\ell+1)}{r^2} + k^2 - V(r)\right) u_\ell = 0.
\end{equation}
The ``centrifugal term'' $\ell(\ell+1)/r^2$ arises from the angular momentum of the $\ell^\text{th}$ partial wave and can be considered an addition to the potential. The discussion so far has considered a single channel, \ie, a single internal state of the atom pair. More generally the wavefunction becomes a vector $\vec{u}_\ell$ with components of each channel, and $V(r)$ is a matrix allowing coupling between the different channels to account for all the possible state pairs of the atoms.

In general, inter-channel coupling allows inelastic collisions where $S$-matrix elements may no-longer have unity modulus. There may also be coupling between different $\ell$. Since we consider elastic collisions with the atomic quantisation axis along the collision axis, we do not have inter-$\ell$ coupling\footnote{Perturbative dipole-dipole interactions, which weakly couple $\ell \pm 2$ can be ignored in the context of this work.}, and we may treat each component separately. Furthermore, threshold laws allow us to ignore large $\ell$ for low energy collisions~\cite{Sadeghpour2000}. For the cold collision experiments considered in this contribution, the scattering is accurately described by $\ell \leq 4$ and the scattering resonances considered are in channels with $\ell = 0$ or $2$. 

\subsection{Jost functions and poles}

As we have seen, the long-range boundary condition of a physical wavefunction is a superposition of incoming and outgoing spherical waves. More rigorously, one can define entire (non-physical) solutions to the radial Schr\"odinger equation with the spherical-wave boundary conditions~\cite{Taylor2006,Sitenko1971,Rakityansky2022}
\begin{equation}\label{eqn:jost_sln}
    \phi_\ell^\pm(k,r) \overset{r\to\infty}{\sim} \exp[\pm\ii (k r - \ell\pi/2)],
\end{equation}
known as the Jost solutions. A physical scattering wavefunction can then be constructed from these
\begin{equation}\label{eqn:scatt_jost}
    u_\ell(E,r) = \frac{\ii}{2}\left[ \mathcal{F}_\ell^\text{in}(E) \phi_\ell^-(k,r) - \mathcal{F}_\ell^\text{out}(E) \phi_\ell^+(k,r)\right],
\end{equation}
introducing the Jost functions $\mathcal{F}^\text{in/out}_\ell(E)$. The $S$ matrix is then defined by the ratio 
\begin{equation}\label{eqn:S_jost}
    S_\ell(E) = \frac{\mathcal{F}_\ell^\text{out}(E)}{\mathcal{F}_\ell^\text{in}(E)}.
\end{equation}
From \eref{eqn:S_jost}, it is apparent that $S_\ell(E)$ will have a pole wherever the Jost function $\mathcal{F}_\ell^\text{in}(E)$ is zero. Furthermore, from \eref{eqn:scatt_jost} and \eref{eqn:jost_sln} one can see that this corresponds to a wavefunction with an exclusively `outgoing' boundary condition, corresponding to the presence of a so-called Gamow or Siegert state~\cite{Gamow1928,Siegert1939} above threshold.

Bound solutions to the Schr\"{o}dinger equation below threshold also coincide with Jost-function zeros. Since $E<0$, $k$ is purely imaginary, and a Jost-function zero with $\imag k > 0$ in \eref{eqn:scatt_jost} ensures that the wavefunction is purely exponentially decreasing as $r\to\infty$. In addition to physically meaningful bound-states with a wavefunction that decays exponentially away in the classically forbidden region, a Jost-function zero may also occur with $\imag k < 0$, corresponding to a non-physical anti-bound state below threshold which is purely exponentially increasing as $r\to\infty$. Despite the non-physical nature of the latter, both bound and anti-bound states can have profound effects on scattering near threshold.

In general, Jost-function zeros and $S$-matrix poles are not found at positive real energies. However, the Jost functions and therefore the $S$ matrix are analytic for well-behaved potentials (except at its poles)~\cite{Taylor2006} and therefore the $S$ matrix on the positive real energy line has an analytic continuation into the complex plane. Poles residing in the complex plane may therefore leave a distinct imprint in the form of resonances for scatting experiments conducted on the experimentally-accessible real energy axis.

\subsection{Resonances and poles}\label{sec:res_poles}

In this work, we consider three different classes of resonances, shown pictorially in \figref{fig:scattering}c.
Firstly, a bound state (dashed blue line) just below the threshold of the entrance channel  strongly affects the near-threshold scattering behaviour. 
Near threshold, atomic interactions are uniquely determined by the scattering length, which is affected by both the long-range potential and the last bound or anti-bound state below threshold. As discussed above, both bound and anti-bound states give rise to Jost-function zeros, and therefore they correspond to $S$-matrix poles below threshold. If the last bound state is close to threshold, the scattering length will be anomalously large. On the other hand, a large negative scattering length instead signals the presence of an anti-bound state, also called a virtual state. The pole associated with either of these is physically inaccessible to scattering experiments, but its presence is visible in the observed scattering just above threshold, resulting in a so-called sub-threshold resonance~\cite{Rakityansky2022}. 

A second class to be found in the entrance channel is the shape resonance. For partial waves with non-zero angular momentum ($\ell > 0$), the centrifugal term of the radial Schr\"odinger equation gives rise to an effective barrier in the potential (solid orange curve in \figref{fig:scattering}c). This barrier can introduce a quasi-bound state above threshold (dashed orange line). An incoming particle matching the energy of the quasi-bound state can resonantly tunnel through the barrier, increasing the duration of the interaction.

The final resonance type considered is the Feshbach resonance, where coupling to a bound state of a closed channel (dashed red line) enhances the interaction. Effectively, the scattering atoms are temporarily bound as a molecule in the closed channel at short range. If the closed channel has a different magnetic moment to the entrance channel, a magnetic field can be used to adjust the position of the resonance relative to the entrance channel. The ability to control atomic interactions with a magnetic field via a Feshbach resonance is a vital tool in ultracold atomic physics~\cite{Chin2010}.

The relationship between $S$-matrix poles and above-threshold resonance phenomena, can be recognized by considering a pole located at a complex energy off the real line, $\bar{E} = E_{\mathrm{r}} - \ii E_\ii$, where $E_{\mathrm{r}}, E_\ii > 0$. Since $\mathcal{F}_\ell^\text{in}(E) = \left[\mathcal{F}_\ell^\text{out}(E^*)\right]^*$~\cite{Rakityansky2022}, then back on the physically meaningful real $E$ axis
the scattering phase is given by [cf. \eref{eq:sell} and \eref{eqn:S_jost}]
\begin{equation}
  \delta_\ell(E) = - \arg\mathcal{F}_\ell^\text{in}(E).
\end{equation}
Expanding $\mathcal{F}_\ell^\text{in}$ about $\bar{E}$ to first order, the phase in this region is
\begin{align}
  \delta_\ell(E)\approx - \arg\left[{\mathcal{F}_\ell^\text{in}}'(\bar{E}) (E - \bar{E})\right]&= - \arg{\mathcal{F}_\ell^\text{in}}'(\bar{E}) - \arg(E - \bar{E}),\nonumber\\
  &= \delta_\text{bg} - \arctan\left(\frac{-\imag \bar{E}}{E - \real \bar{E}}\right). \label{eqn:pole_res}
\end{align}
The second term, the phase winding associated with being in the vicinity of a pole, is the source of resonant behaviour while the background phase shift $\delta_(\rm bg)$ encapsulates the remaining non-resonant scattering behaviour.
In particular, \eref{eqn:pole_res} replicates the general form for the scattering phase near a Feshbach resonance along the real energy line~\cite{Chin2010}:
\begin{equation}\label{eqn:energy_res}
    \delta_\ell(E) = \delta_\text{bg} - \arctan\left(\frac{\Gamma/2}{E - E_C - \delta E}\right),
\end{equation}
where $\delta_\text{bg}$, the width $\Gamma$ and shift $\delta E$ all generally change with energy, associated with a bound-state in a closed channel at energy $E_C$. The coupling between open and closed channels causes a resonance at $E_\text{res} = E_C - \delta E$. By inspection of equations \eref{eqn:pole_res} and \eref{eqn:energy_res}, we see that the resonance is equivalent to a pole at $\bar{E} = E_\text{res} - \ii\Gamma/2$.

In the present study, we consider magnetically-tunable resonances from two different perspectives: `viewing' them in either energy or magnetic field. Above, we considered how a resonance is visible as a function of collision energy, in the phase winding while moving past a static resonance pole (i.e. at a constant magnetic field).
Alternatively, one may fix the collision energy and observe a resonance profile in magnetic field, tuning the resonance pole across the chosen energy. In the latter case, the resonance profile observed in magnetic field $B$ is described by the Breit-Wigner profile~\cite{Chilcott2022},
\begin{equation}\label{eqn:mag_res}
    \delta_\ell(B) = \delta_\text{bg} + \arctan\left(\frac{\Gamma_B/2}{B - B_\text{res}}\right),
\end{equation}
where the width $\Gamma_B$ and position $B_\text{res}$ of the resonance will depend upon energy.

Our discussion of the $S$ matrix and the scattering phase is connected to experimental observations by noting that the scattering phase determines the partial scattering cross-section~\cite{Friedrich2013},
\begin{equation}
    \sigma_\ell = \frac{4\pi(2\ell+1)g}{k^2} \sin^2\delta_\ell,
\end{equation}
where the factor $g$ is included to account for collisions of indistinguishable particles. For distinguishable particles (e.g., between \K{40} and \Rb{87}, or \Rb{87} prepared in different internal quantum states~\cite{Mellish2007}), $g = 1$. For situations instead considering indistinguishable bosons (fermions), $g = 2$ when $\ell$ is even (odd) and $g = 0$ otherwise. The total scattering cross-section is given by the sum of the partial cross-section, though as previously mentioned, the experiments considered here are accurately described by the terms with $\ell \leq 4$.

\subsection{A simple conceptual model for pole interactions}\label{sec:concept}

\begin{table*}
  \centering
  \renewcommand{\arraystretch}{1.3}
  \caption{\label{tbl:pclass}Classifications of the coupled two-channel model~\cite{Chilcott2021}.}
  \begin{tabular}{|p{0.2\linewidth}|c|c|c|}
    \hline
    \textbf{Case} & \textbf{I} & \textbf{II} & \textbf{III}\\
    \hline
    $(\epsilon_1 - \epsilon_2)^2 + 4\omega^2$ & Positive & Zero & Negative\\
    \hline
    Coupling Strength  & $4|\omega| < |\gamma_1 - \gamma_2|$ & $4|\omega| = |\gamma_1 - \gamma_2|$ & $4|\omega| > |\gamma_1 - \gamma_2|$ \\
    \hline
    \multirow{2}{*}{Crossing} &  Real Energy & Poles Coincide & Imaginary Energy\\
    & $\real \mathcal{E}_1(B_0) = \real \mathcal{E}_2(B_0)$ & $\mathcal{E}_1(B_0) = \mathcal{E}_2(B_0)$ & $\imag \mathcal{E}_1(B_0) = \imag \mathcal{E}_2(B_0)$\\
    \hline
    {Example pole trajectories. \vspace{-0.5em} \newline  \hspace*{3em} \includegraphics[width=0.38\linewidth]{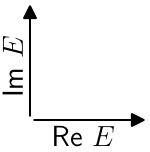}}
    & \raisebox{-4.5em}{\includegraphics[width=0.22\linewidth]{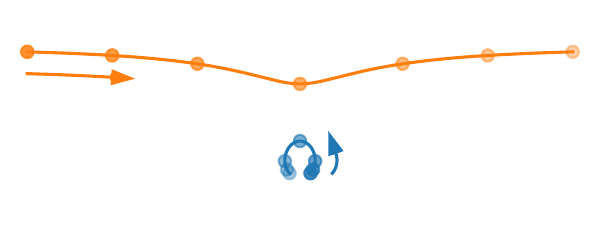}} & \raisebox{-4.5em}{\includegraphics[width=0.22\linewidth]{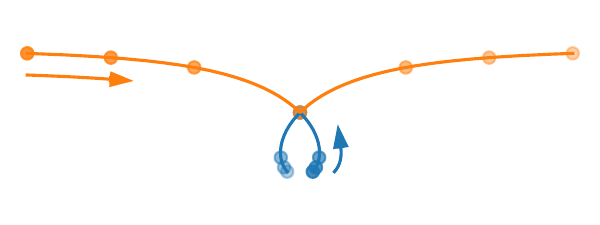}} & \raisebox{-4.5em}{\includegraphics[width=0.22\linewidth]{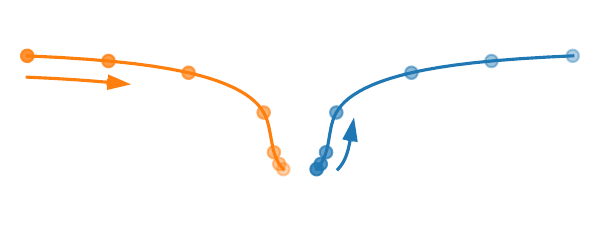}}\\
    \hline
  \end{tabular}  
\end{table*}

To form a simplified model for the interplay of two $S$-matrix poles, we treat the resonance poles above threshold as complex eigenvalues of a non-Hermitian Hamiltonian, where the imaginary components represent the decay from the quasi-bound states into the scattering continuum. In the case of Feshbach resonances, one may obtain equivalent models by projecting the system Hamiltonian onto the closed-channel subspace~\cite{Feshbach1958,Feshbach1962}. Alternatively, one could employ quantum defect theory to separate out the long-range behaviour of the open channel from the short-range inter-channel coupling~\cite{Mies1984,Julienne2006,Chilcott2022,Naidon2024} of the closed channel resonances. For our purpose, however, we also need to explicitly include resonances of the open channel.

Consider two non-interacting $S$-matrix poles located at complex energies $\varepsilon_n(B) = E_n(B) - \frac{\ii}{2}\gamma_n$. The interaction between these with (real) strength $\omega$ can be described by the effective Hamiltonian over the non-interacting poles~\cite{Rotter2001,Okoowicz2003}
\begin{equation}\label{eqn:simple_model}
  H = \begin{bmatrix}\varepsilon_1(B) & \omega \\ \omega & \varepsilon_2(B)\end{bmatrix}.
\end{equation}
With the coupling added, the positions of the poles are given by the eigenvalues of \eref{eqn:simple_model},
\begin{equation}\label{eqn:model_eig}
  \mathcal{E}_\pm = \frac{\varepsilon_1 + \varepsilon_2}{2} \pm \frac{1}{2}
  \sqrt{(\varepsilon_1 - \varepsilon_2)^2 + 4 \omega^2}.
\end{equation}

For simplicity, we consider one pole to have a fixed uncoupled position while the real energy component of the other pole increases linearly with $B$, and that at $B = B_0$ the real parts of the energies $\varepsilon_1$ and $\varepsilon_2$ coincide. When the coupling is introduced, the poles cross in  different ways, as summarised in \tref{tbl:pclass}. The crossing behaviour can be classified into three characteristic cases~\cite{Okoowicz2003} delineated by the argument of the square-root in \eref{eqn:model_eig} at $B = B_0$. Specifically, if it evaluates as zero ($|\gamma_1 - \gamma_2| = 4|\omega|$), positive ($|\gamma_1 - \gamma_2| > 4|\omega|$), or negative ($|\gamma_1 - \gamma_2| < 4|\omega|$).
When $|\omega|=|\gamma_1-\gamma_2|/4$ (case II), the two
states and the corresponding poles will coalesce exactly ($\mathcal{E}_1=\mathcal{E}_2 $) at $B =
B_0$, producing a so-called exceptional point
\cite{Kato1966,Heiss2012}. In the other two cases, only the real (case I) or imaginary (case III) components of the poles coincide at $B_0$. As elucidated graphically in \tref{tbl:pclass}, the pole trajectory for case I shows the two poles to be pulled together as they cross in real energy. In case III, one pole is pushed away by the approach of the other and they do not cross. 
}
\section{Scattering calculations and pole positions}

We ascertain the position of the poles by analytically continuing a calculated $S$ matrix into the complex energy plane as in Ref.~\cite{Chilcott2021}. This process starts with $S$-matrix elements, calculated at positive real energies using coupled-channels calculations. To these physically-meaningful solutions, we fit a Pad\'e approximant (using a linear least-squares approach) whose domain extends into the complex energy plane. In general, the Pad\'e approximant $f^{[N,M]}(z)$ of the function $f$ over the complex variable $z$, is defined by
\begin{equation}
  f^{[N,M]}(z) = \frac{P(z)}{Q(z)},
\end{equation}
where $P(z)$ and $Q(z)$ are polynomials of degree $N$, and $M$
respectively, with $M=N=4$ in the computations of this work. The poles of the approximant are trivially extracted as the roots of the polynomial $Q$. $S$-matrix poles are calculated for a range of magnetic fields, and the trajectories stitched together by increasing the field and taking the closest calculated pole as the next position. The extracted poles can be sensitive to numerical noise in the $S$ matrix, so non-physical pole jumps are filtered out and the trajectory is smoothed. 
Pad\'e approximants are particularly useful for the analytic continuation of calculations which are limited to a certain domain~\cite{Vidberg1977}. However, there are alternative approaches to this problem which are less susceptible to error/noise~\cite{Kaufmann2023}, but these do not allow such a simple extraction of the pole positions.

\section{Interactions between single-channel and Feshbach resonances}
\begin{figure}
  \includegraphics[width=\textwidth]{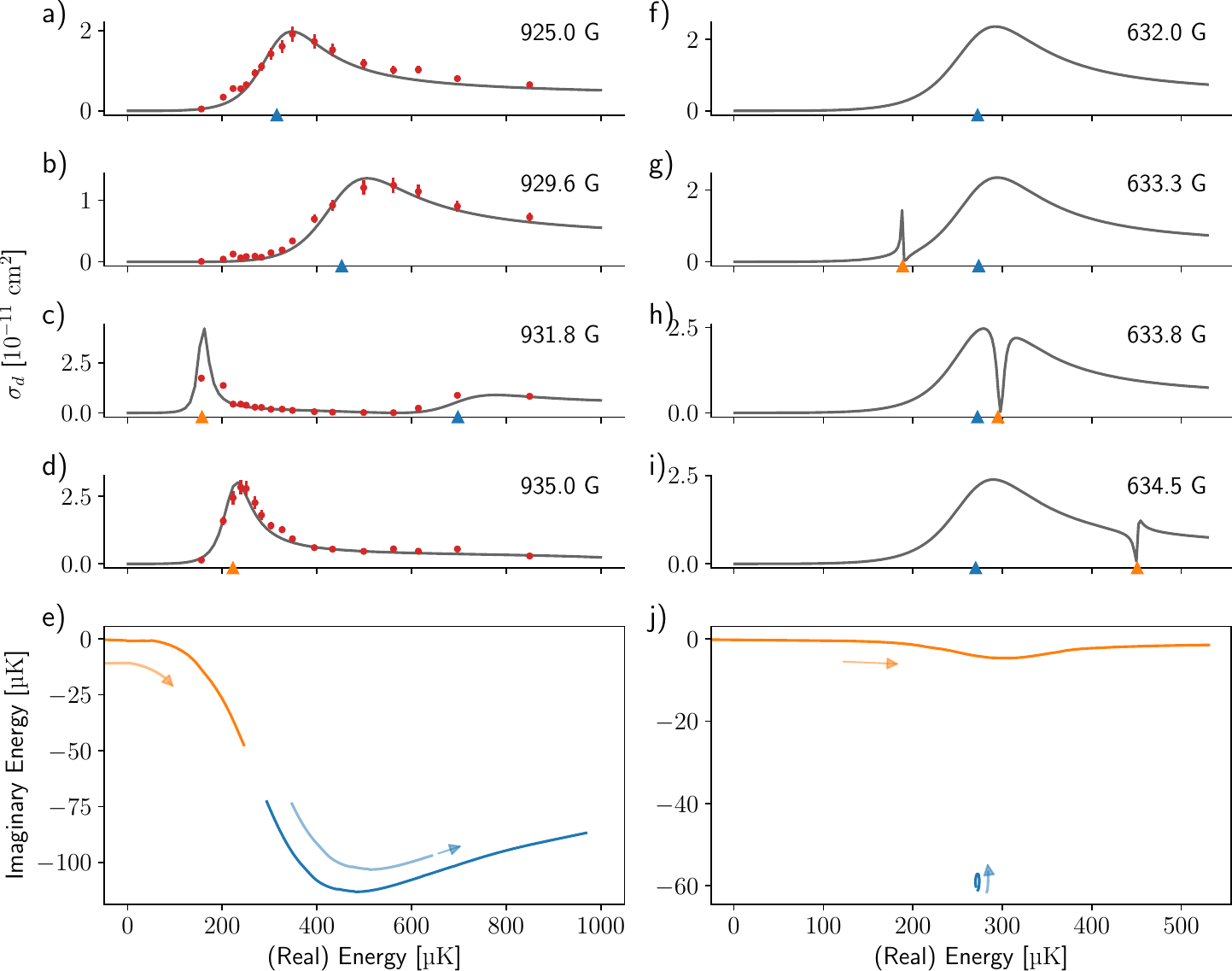}
\caption{The shape resonance pole of \Rb{87} performing two distinct dances with different Feshbach resonances: one near \SI{930}{\gauss} in the left column, and \SI{632}{\gauss} on the right. The upper panels (a,b,c,d and f,g,h,i) show the calculated $d$-wave cross-section for several magnetic fields, with increasing field moving up the page. The red dots show measured values for the \SI{930}{\gauss} resonance. At the bottom of pane, colour-coded triangles specify the real-energy position of the poles at that field. The complex energy position of the poles themselves are shown in the lower panes (e,j), with the trajectories evolving in the direction of the arrow with increasing magnetic field. The magnetic field tunes the position of the Feshbach resonance, where the \SI{930}{\gauss} resonance avoids crossing in real energy, and the \SI{632}{\gauss} resonance avoids crossing in imaginary energy.}
\label{fig:ShapePoleDances}       
\end{figure}

\todo{Section \ref{sec:res_poles} introduced three classes of resonances which all can be associated with $S$-matrix poles. Two of these, shape resonances and sub-threshold resonances, are both features of a single open channel---the entrance channel. In the following, we consider the interaction of such single-channel resonances with a Feshbach resonance hosted in a closed channel with coupling to the entrance channel.}

\subsection{Feshbach and shape resonance interactions.}\label{sec:feshbach-shape}
The $d$-wave ($\ell = 2$) potential of \Rb{87} hosts a prominent shape resonance situated near \SI{300}{\micro\kelvin}. This resonance is visible as the dominant cross-section peak in Figs.~\ref{fig:ShapePoleDances}a,f (black curve). With atoms in the \ket{F=1, m_F=1} state, we study this resonance's interactions with two particular $d$-wave Feshbach resonances at 632~G and 930~G, respectively---here, the magnetic field values refer to where the Feshbach resonances cross threshold. These resonances were selected to elucidate and exemplify cases I and III of \tref{tbl:pclass}.

The left column of \figref{fig:ShapePoleDances} describes the \SI{930}{\gauss} resonance, for which we have considered the pole trajectories previously~\cite{Chilcott2021} and review here. Figures~\ref{fig:ShapePoleDances}a-d present the predicted (line) and measured (red dots) $d$-wave scattering cross-section as measured for select fields around \SI{930}{\gauss}, In particular, it is apparent how the \SI{\sim 300}{\micro\kelvin} shape resonance moves up in energy and that this movement begins even before the Feshbach resonance reaches threshold. After the Feshbach resonance pole crosses threshold, it replaces the shape resonance near \SI{300}{\micro\kelvin}. This behaviour is highlighted in Figs.~\ref{fig:ShapePoleDances}a-d by the colour-coded triangles highlighting the real component of the resonance poles. The complete trajectory of the poles, \figref{fig:ShapePoleDances}e, shows the poles in the complex energy plane with the arrows showing the direction of pole motion with increasing field. This exposes the effectively repulsive interaction of the poles, akin to an avoided crossing of class III described above (cf. \tref{tbl:pclass}).

\begin{figure}
    \includegraphics[width=\textwidth]{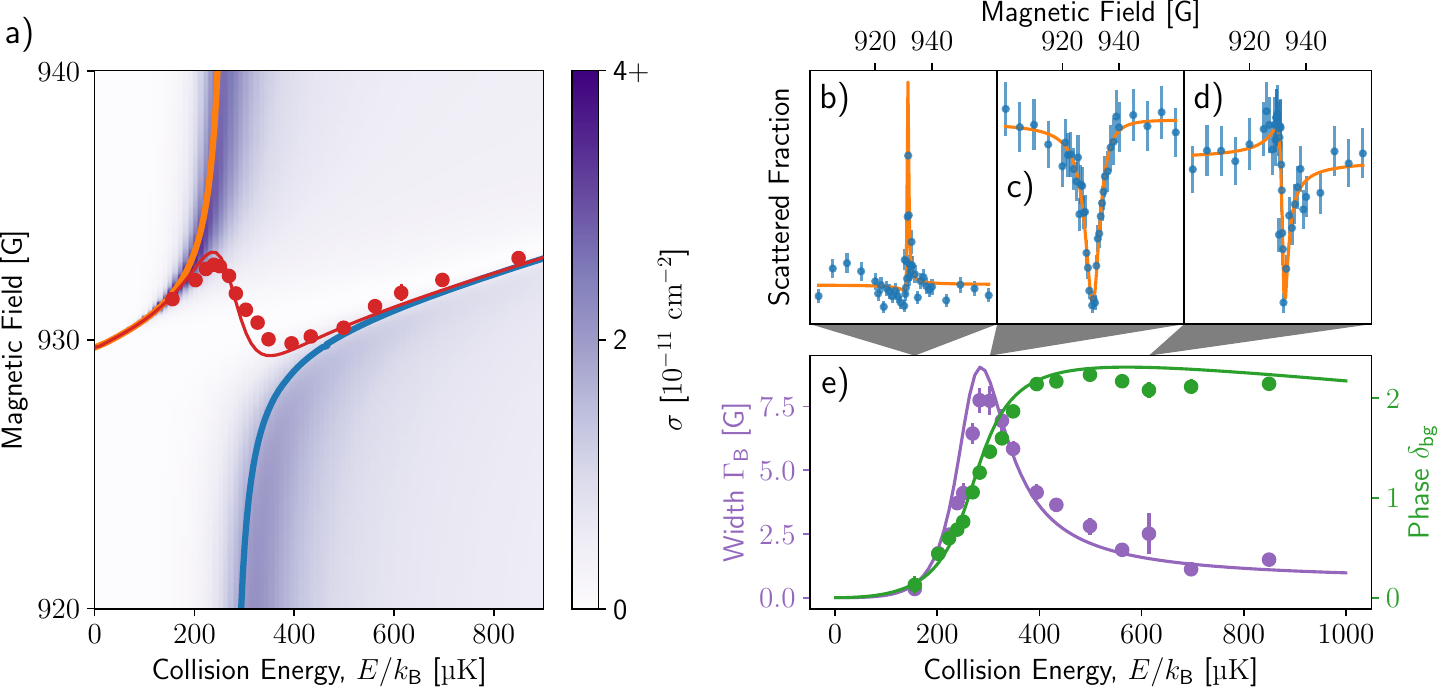}
    \caption{Scattering near two resonances. a) The observed (red dots) and predicted (red line) position of the Feshbach resonance vieved in magnetic field. The orange and blue curves are the real energy component of the corresponding poles in \figref{fig:ShapePoleDances}e. These are overlaid on calculations of the scattering cross-section. b,c,d) Observations of atomic scattering strength at fixed energies. e) The width and backgound phase of the resonance observations in magnetic field.}
    \label{fig:930data}
\end{figure}

\Figref{fig:930data} shows properties of the experimentally measured Fano profiles `viewed in magnetic field'. By this, we mean an observation of the resonant feature, where the magnetic field is scanned while the collision energy is fixed. This highlights some interesting differences between viewing the resonance phenomenon in energy or magnetic field. Consider \figref{fig:930data}a, which shows the real-energy trajectories of the two poles (orange, blue) as well as the observed (red dots) and predicted (red line) position of the resonance feature (modelled by a Fano profile) when viewed in magnetic field. The poles undergo an avoided crossing while the magnetic position swaps between the two: in the low and high collision energy limits, the magnetic observation and the energy observation places the resonance at the same location in $E-B$ space. In the crossing region, the magnetic position swaps between the two poles, and both poles are involved in producing the magnetic resonance profile. \Figref{fig:930data}e shows that as the magnetic position swaps, the background phase and therefore shape of the magnetic resonance changes, indicated by the experimental observations and fitted Fano profiles in Figs.~\ref{fig:930data}b-d. Additionally, the width of the magnetic resonance feature takes its maximum value at the position of the avoided crossing.

In stark contrast to the strong interactions of the \SI{930}{\gauss} resonance, the right column of \figref{fig:ShapePoleDances} shows that the \SI{632}{\gauss} resonance barely perturbs the shape resonance and simply passes over it. This is especially evident from the lack of movement of the shape resonance pole in \figref{fig:ShapePoleDances}j. The cross-sections, Figs.~\ref{fig:ShapePoleDances}f-i, show that as the Feshbach resonance passes the shape resonance, it changes from a scattering enhancement to a suppression and appears to split the shape resonance in two as it does so. The pole trajectory corresponds to the case I interaction of \tref{tbl:pclass}.
The interaction of this Feshbach-shape-resonance pair is the subject of Ref~\cite{Drr2005}, where the authors discuss the role of the shape resonance in the dissociation of Feshbach molecules quasi-bound by this resonance, from the experiments of Ref.~\cite{Volz2005}.

\todo{Above, we have contrasted two interactions of Feshbach and shape resonances with markedly different behaviour observable in collisions at physical energies. In both cases, the resonance interplay is manifest in the crossing trajectories of their $S$-matrix poles, which are of different classes captured by the simple non-Hermitian model of \sref{sec:concept}.}

\subsection{Feshbach and sub-threshold resonances}
\begin{figure}
    \includegraphics[width=\textwidth]{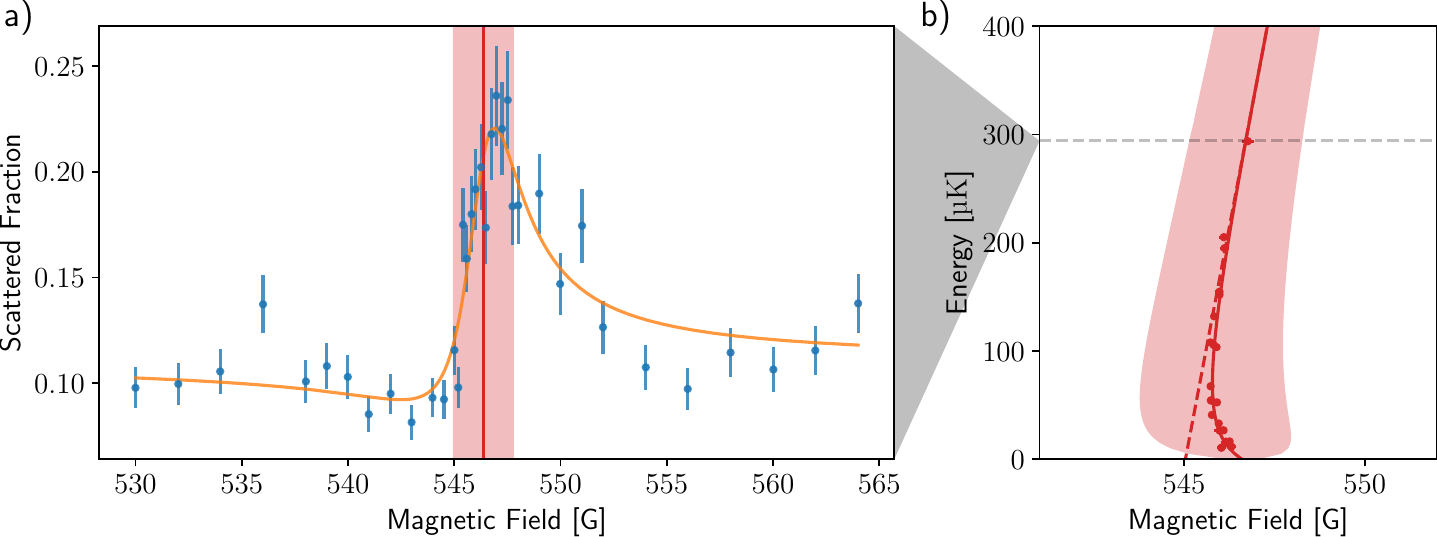}
    \caption{a) The measured scattered fraction of \K{40}-\Rb{87} collisions displaying a Beutler-Fano profile as a function of $B$-field with the collision energy fixed at $E=\SI{294}{\micro\kelvin}$. The red vertical line is the inferred Fano profile position~\cite{Thomas2018} $B_\text{res}$ while its width $\Gamma_\text{B}$ is represented by the shaded region. b) The evolution of the position and width of the Beutler-Fano profile with collision energy. Measurements of $B_\text{res}$ are shown as dots, with error-bars smaller than the markers. The linear tuning of the closed channel is shown as a dashed line. Near threshold the actual position curves away from this due to interaction with the virtual state below threshold of the open channel. The experimental data are from Ref.~\cite{Thomas2018}.}
    \label{fig:KRbFano}
\end{figure}

The \K{40}-\Rb{87} pair in the \ket{F,m_F} hyperfine states \ket{9/2,-9/2} and \ket{1,1}, respectively, has a negative scattering length of approximately $-185a_0$~\cite{Klempt2008}. This results from  the presence of an anti-bound (virtual) state just below threshold. \Figref{fig:KRbFano}a shows the result of colliding atoms at an energy of \SI{\sim 293}{\micro\kelvin}, revealing a Beutler-Fano profile when scanning the magnetic field from \SI{530}{\gauss} to \SI{565}{\gauss}.  The Fano profile results from a magnetic Feshbach resonance, and its position will shift as the collision energy is lowered (\figref{fig:KRbFano}b). Far above threshold the behaviour is captured by a single isolated resonance pole moving on a straight line in $E$-$B$ parameter space. However, as the Feshbach resonance pole is tuned towards threshold from above, the interaction with the sub-threshold resonance---the anti-bound state---will make an imprint on the position of the recorded Fano profile as the $S$-matrix poles flow. Intriguingly, the trajectory of the Fano profile position will depend crucially on the anti-bound nature of the sub-threshold pole~\cite{Marcelis2004}. As such the pole flow and its scattering imprint reveals the sign of the scattering length -- something a simple low-energy cross section measurement cannot do. In Ref.~\cite{Thomas2018} we reported on the experimental observation of the non-monotone trajectory of the Fano profile position as a Feshbach resonance was tuned towards threshold from above, and we revisit this data here.

\begin{figure}
      \includegraphics[width=\textwidth]{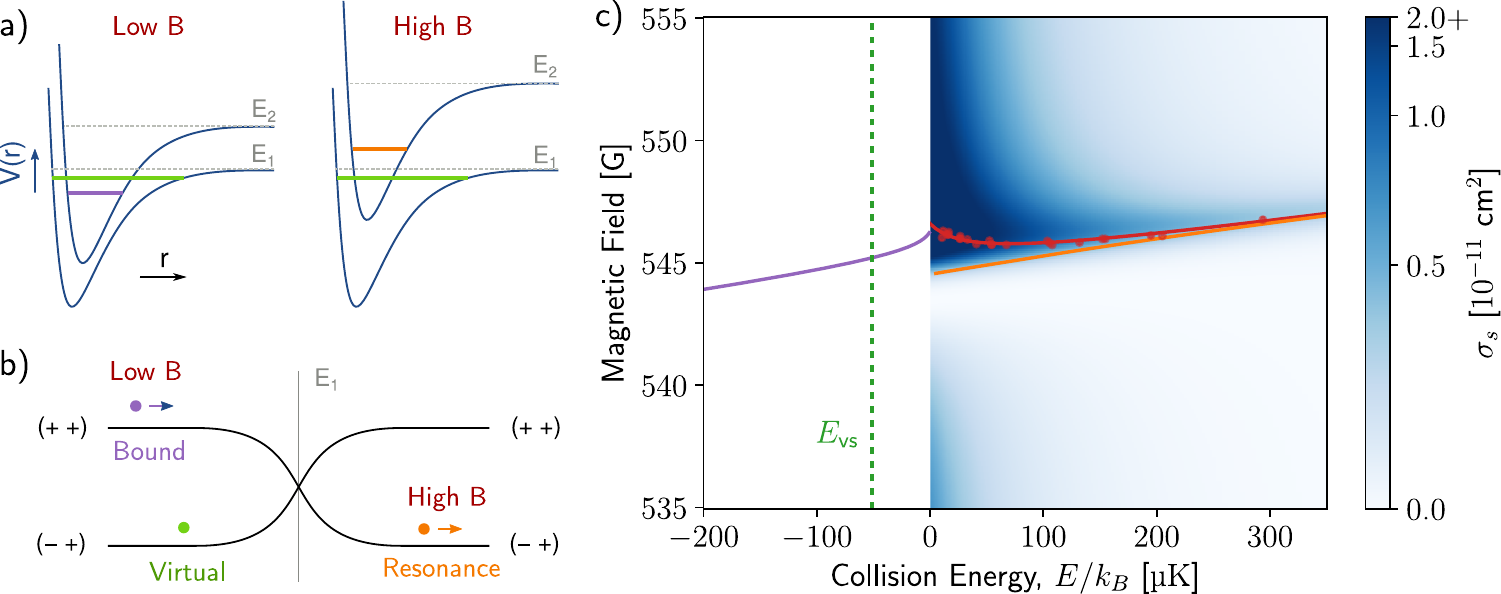}
      \caption{Scattering of \K{40} with \Rb{87} in the vicinity of a Feshbach and virtual-state induced resonance.
      a) At low $B$-fields and collision energies of $E_1 < E < E_2$, the interaction potentials shown are open ($E_1$) and closed ($E_2$), hosting virtual states (green) and bound states (purple) respectively. At higher magnetic fields, the closed channel is tuned with respect to the entrance channel, such that the previously-bound state becomes quasi-bound---a Feshbach resonance accessible to scattering in the open channel (orange).
      b) The Riemann sheets on which the poles lie, at fields much higher and lower than $B_\text{res}(E=0)$. The sheets are labelled with the signs of $\imag k$ in the two channels. The bound and resonant poles correspond to the states in the closed channel potentials above, and the arrows indicate their motion with increasing magnetic field.
      c) The measured (red dots) and calculated (red line) position $B_\text{res}$ of the resonance observed in magnetic field, overlaid the calculated scattering cross section. The real part of the energy for the $S$-matrix resonance pole extracted using a Pad\'{e} approximant is shown in orange. The position of the Feshbach bound-state below threshold is shown in purple, and the position of the virtual state in green.}
      \label{fig:KRbPoles}
\end{figure}

\Figref{fig:KRbPoles}a shows schematic interatomic potentials which, for collision energies of $E_1 < E < E_2$, correspond to an open (entrance) channel hosting a anti-bound state (green line) and a closed channel that, depending on the $B$-field, will host a bound (purple line) or quasi-bound state (orange line)---a Feshbach resonance. In general, we measure energy with respect to the entrance channel threshold---that is, we have defined $E = 0$ at $E_1$ and the low energy collisions we consider have $E \ll E_2$. At low magnetic fields, the closed channel bound state is below the open channel threshold, and it therefore gives rise to an $S$-matrix pole in the entrance channel on the negative real energy axis of the $(++)$ Riemann sheet. Meanwhile, a pole corresponding to the anti-bound state is located on the negative real energy axis of the $(-+)$ Riemann sheet. The locations of the two poles on separate Riemann sheets are illustrated in \figref{fig:KRbPoles}b. At high magnetic fields, the bound state of the closed channel is brought above the open channel threshold, where it becomes quasi-bound and forms a resonant state. During the course of this, the associated $S$-matrix pole moves onto the $(-+)$ Riemann sheet, off (and below) the real positive energy axis.
Since $E \propto k^2$, \ $S(E)$ needs to be specified on a domain of  Riemann sheets. These can be designated according to the sign of $\imag k$, and in a system with more than one channel, there is hence a sheet for all possible combinations of the signs for $\imag k_i$ in each channel. At threshold of the entrance channel, $E_1$, the two sheets have a branch point, and they have a branch cut along the real energy line for $E > E_1$. For the two channel scattering problem, the sheets are labelled by a pair of signs~\cite{Rakityansky2022}, introducing a total of four sheets. \Figref{fig:KRbPoles}b shows the two relevant sheets that come into play and presents the pole locations at the two magnetic field extremes---low and high---corresponding to those in \figref{fig:KRbPoles}a.

The region between the two magnetic field extremes of \figref{fig:KRbPoles}b is explored in \figref{fig:KRbPoles}c, which shows the magnetic position of the Fano profile, as it curves near threshold (red line and points). Meeting threshold at $B_\text{res}(E=0) = \SI{546.606(22)}{\gauss}$~\cite{Thomas2017}, the magnetic position connects with the bound-state position produced by coupled-channels calculations (purple line). The position of the virtual state is highlighted (green dashed line) extracted from the near-threshold scattering phase far from the presence of the bound-state as $\delta(k) = -k a_\text{bg} - \arctan(k/\kappa_\text{vs})$ where the virtual-state energy is $E_\text{vs} = -\hbar^2\kappa_\text{vs}^2/(2\mu)$~\cite{Marcelis2004}. The real-energy position of the resonance pole above threshold, extracted by the Pad\'e approximant method is shown in orange. At higher energies the magnetic resonance position and the resonance pole coincide, while approaching threshold the two diverge: the bound-state and the resonance pole do not coincide, indicating that the bound-state pole does not smoothly transition to become the resonance pole.

As demonstrated for the shape-Feshbach-resonance interaction in \sref{sec:feshbach-shape}, the divergence of the magnetic resonance position is an indication of multiple poles at play. At fields just above \SI{547}{\gauss}, where the bound-state pole below threshold has disappeared, the cross-section near threshold is anomalously large, yet the resonance pole is far away, suggesting a below-threshold pole higher than the nominal position of the anti-bound state. Bortolotti \etal~\cite{Bortolotti2008} predicts the emergence of a non-physical pole for $B>B_\text{res}$, which might serve to explain the strong physical effect we witness here. Our Pad\'e approximant technique is not adequate for locating poles below threshold and critically does not ascribe physical meaning to such poles. Hence further studies based on alternative approaches would be warranted to elucidate the detailed pole dynamics at threshold. 

\section{Interaction between two Feshbach resonances}

When considering two interacting Feshbach resonances, a particularly interesting possibility arises: the coupling between two Feshbach resonances and the open channel can cause the coupling between the open channel and one resonance state to disappear. The decoupling means that the resonance state can no longer decay and the result is a bound state suspended above threshold. This flavour of a ``bound state in the continuum'' (BIC) was predicted by Friedrich and Wintgen~\cite{Friedrich1985}, after the effect was almost observed~\cite{Neukammer1985}; the experiment was restricted to a discrete parameter space so that the BIC conditions could not be truly satisfied.

\begin{figure}
    \centering
    \includegraphics[width=0.5\linewidth]{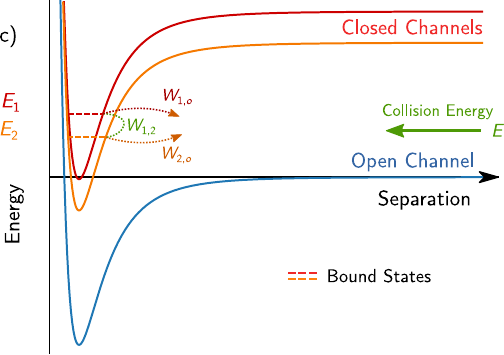}
    \caption{A schematic view of two Feshbach resonances interacting with the open channel. The bound states (dashed lines) in the closed channels (orange, red) are located at energies $E_i$, and possess interchannel couplings $W_{i,j}$. In the system presented, the two closed channels tune in magnetic field at different rates so that $E_1$ and $E_2$ cross near $E/k_\text{B} = \SI{1770}{\micro\kelvin}$.}
    \label{fig:BIC_potentials}
\end{figure}

\todo{While originally transpiring from quantum mechanics~\cite{Neuman1929}, BICs are a general wave phenomenon observed in a variety of settings~\cite{Hsu2016} from acoustic resonators~\cite{Huang2022} to plasmon systems~\cite{Azzam2018}. They are of particular interest in photonic resonators~\cite{Capasso1992,Plotnik2011,Hsu2013,Hu2022} where the massive $Q$ factor they effect is expected to have applications in lasing, non-linear optics and sensing~\cite{Sadreev2021}.
In atomic scattering, a BIC has been proposed to provide an efficient pathway for the production of Feshbach molecules at energies above threshold~\cite{BDeb2014}. Specifically, Ref.~\cite{BDeb2014} considers a BIC in the scattering continuum induced by laser-coupling via photoassociation resonances. The possibility of coupling states through added external electromagnetic fields offers a flexible way to engineer BICs between two atoms. BICs, however, may also emerge through the inherent hyperfine coupling between atomic states if their relative locations can be tuned. As an outlook towards future experiments with our laser-based atom-collider, we consider the interaction of two magnetically tunable Feshbach resonances with different tuning rates. As shown below, a BIC may form around the (avoided) crossing of the two coupled resonances at a particular collision energy and magnetic field.}

\todo{\Figref{fig:BIC_potentials} shows the bound-state energies $E_{i}$ and couplings $W_{i,j}$ for a pair of Feshbach resonances in two closed channels ($i=1,2$) interacting with the open entrance channel ($i=o$). A simplified model~\cite{Friedrich1985,Friedrich2013} establishes that a BIC will be formed in this scenario if
\begin{equation}\label{eqn:bic_criterion}
    E_1 - E_2 = W_{2,1}\frac{W_{1,o}}{W_{2,o}} - W_{1,2}\frac{W_{2,o}}{W_{1,o}}.
\end{equation}
In a continuous tuning space, this criterion can be met for any pair of closed channel resonances at some collision energy $E$ where
\begin{equation}
    E = E_1 - W_{2,1}\frac{W_{1,o}}{W_{2,o}} = E_2 - W_{1,2}\frac{W_{2,o}}{W_{1,o}}.
\end{equation}

\begin{figure}
      \includegraphics[width=\textwidth]{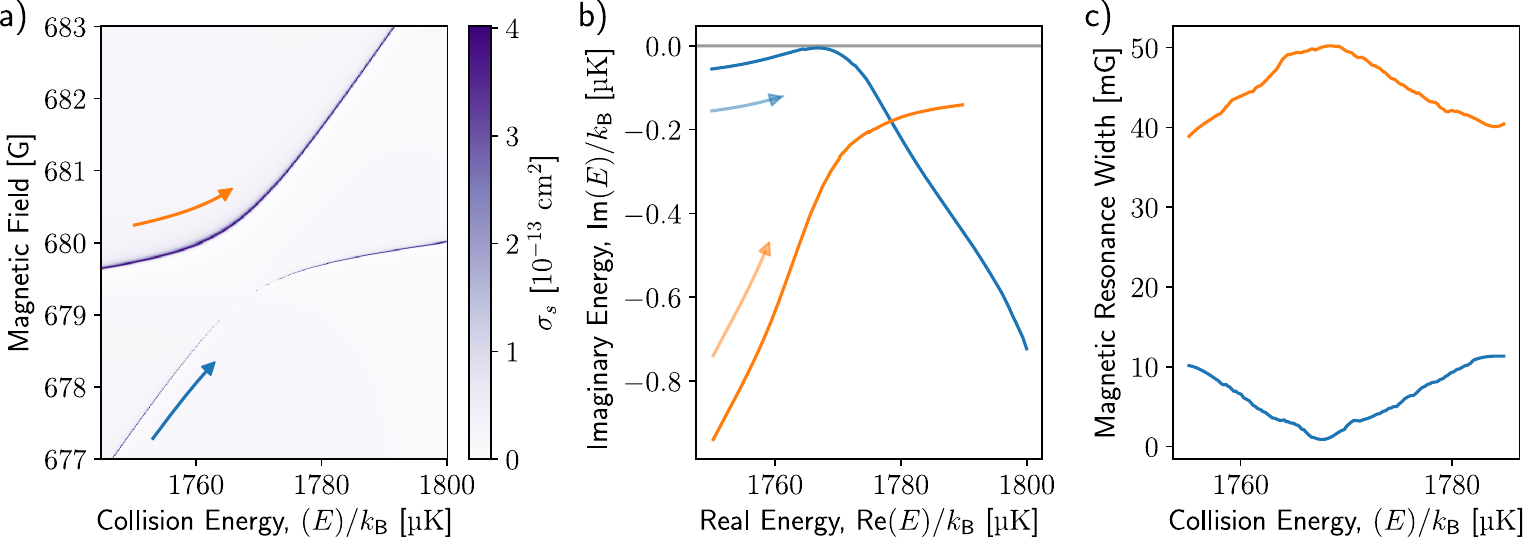}
      \caption{a) The scattering behaviour during the avoided crossing of two Feshbach resonances, noting that the width of the lower resonance vanishes. b) The two poles associated with the Feshbach resonances, noting that the pole associated with the narrowing resonance approaches the real energy line, representing a bound state in the continuum (BIC). In both plots, the colour-coded arrows demonstrate the direction of the resonance with increasing magnetic field. c) The magnetic width $\Gamma_B$ of the two Feshbach resonances, one of which also vanishes at the BIC.}
      \label{fig:bic_poles}
\end{figure}

In \figref{fig:bic_poles} we consider $s$-wave scattering of \Rb{87} atoms in the \ket{F=1, m_F=0} state. At collision energies of around \SI{1770}{\micro\kelvin} and fields of around \SI{680}{\gauss}, two Feshbach resonances cross. The Feshbach resonances labelled blue and orange cross threshold at \SI{414}{\gauss} and \SI{662}{\gauss}, respectively.
In their interaction, a BIC emerges, visible as a narrowing and disappearance of the scattering peak of the blue-labelled resonance during their avoided crossing. In the pole flow of \figref{fig:bic_poles}b, we observe that the pole associated with this resonance approaches the real energy line. At the real energy line, the energy width vanishes---the hallmark of BIC formation. This is also captured in \figref{fig:bic_poles}c where the magnetic width likewise disappears at the BIC.

Our analysis of \eref{eqn:simple_model} assumed that one resonance remained stationary. We now have both resonances tuning with magnetic field at different rates. Accounting for the simultaneous tuning, the interaction between these two poles can be classed as a case III interaction from \tref{tbl:pclass}, similar to the \SI{930}{\gauss} resonance of \sref{sec:feshbach-shape}. In particular, the poles change which Feshbach resonance they represent, consistent with the visible avoided crossing in the scattering cross-section, c.f. \figref{fig:930data}. 

Extending \eref{eqn:simple_model} to describe a BIC, we introduce the role of interference via the open channel with a complex coupling~\cite{Devdariani1976,Hsu2016}, replacing the real inter-channel coupling $\omega$ with $\bar{\omega} = \omega - {\ii\sqrt{\gamma_1\gamma_2}}/2$:
\begin{equation}\label{eqn:simple_model2}
  H_{\rm BIC} = \begin{bmatrix}E_1(B) & \omega \\ \omega & E_2(B)\end{bmatrix}-\frac{\ii}{2}\begin{bmatrix}\gamma_1 & \sqrt{\gamma_1\gamma_2} \\ \sqrt{\gamma_1\gamma_2} & \gamma_2\end{bmatrix}.
\end{equation}
The updated Hamiltonian now has one real eigenvalue (a BIC) when $\epsilon_1 - \epsilon_2 = \omega(\gamma_1 - \gamma_2)/\sqrt{\gamma_1 \gamma_2}$, corresponding precisely to \eref{eqn:bic_criterion}.

Which pole reaches the real line and turns into a BIC is decided by the relative coupling strengths and the sign of the coupling product $W_{2,1} W_{1,o} W_{2,o}$~\cite{Friedrich1985}. For our updated conceptual model we have implicitly chosen $\gamma_i > 0$ but we can allow the real component $\omega$ of the coupling strength to be negative to account for the two possible overall signs of the coupling product. Assuming that $\gamma_1 \neq \gamma_2$, the model indicates that the initially narrower resonance \todo{(smallest $\gamma_i$)} will become the BIC, \todo{except in the case of} strong coupling ($4|\omega| > |\gamma_1 - \gamma_2|$) with $\omega < 0$, where the poles cross sufficiently for the wider resonance to form a BIC. 
For critical coupling ($4|\omega| = |\gamma_1 - \gamma_2|$), there is an exceptional point above (below) the BIC for positive (negative) $\omega$.
In the case that $\gamma_1 = \gamma_2$, the pole with lower (higher) real energy will form a BIC for positive (negative) $\omega$.

The BIC model of Ref.~\cite{Friedrich1985} states that the sum of the (energy) widths of the resonances remains approximately constant during the interaction, with the narrowing of a forming BIC necessitating that the other resonance broadens. Similarly, the sum of the imaginary components of the eigenvalues of \eref{eqn:simple_model2} is constant: $\imag (\mathcal{E}_+ + \mathcal{E}_-) = -(\gamma_1 + \gamma_2)/2$. \Figref{fig:bic_poles}b shows that this is not true for the calculated poles of the physical system we study. Rather \figref{fig:bic_poles}c indicates that the sum of the two magnetic widths is instead approximately constant. The ratio between the energy and magnetic widths for each resonance is given by the rate at which the resonance tunes ($\dd \varepsilon_i / \dd B$). From \fref{fig:bic_poles}a it is apparent that this ratio is not constant.}

\todo{The question remains if the BIC in \fref{fig:bic_poles} is observable in our collider, e.g., as a vanishing resonant scattering feature in the lower branch of \fref{fig:bic_poles}a. Unfortunately, as described in Ref.~\cite{Horvath2017} the use of sub-microkelvin cold clouds does not necessarily provide sub-microkelvin energy resolution in the laser-based scheme. Rather, the energy spread in an experiment colliding clouds at an energy $E_{\rm nom}$ is $\text{var}(E)\approx\sqrt{2E_{\rm nom}k_B T}$, where $T$ is the cloud temperature. This is so, because the laser-based scheme adds the same velocity to each particle in an accelerating projectile ensemble rather than the same energy as a conventional particle accelerator would do \cite{Reiser1994}. As a result, even 200~nK cold clouds would lead to a $\sim$\SI{27}{\micro\kelvin} energy spread in an $E_{\rm nom}/k_{\rm B}=$\SI{1770}{\micro\kelvin} collision experiment. Observations of this BIC would then rely on energy-broadened observations of narrow scattering features.}

\section{Summary and discussion}
\todo{
The experiments and computations presented above expound the $S$-matrix pole behaviours in resonance interactions of atomic collisions. In particular, the physical manifestations of the pole interplay can be seen in the parameter spaces of energy and magnetic field, accessible to an optical collider manipulating samples of ultracold atoms.

We have identified realisations of different classes of interactions between resonances. For Feshbach and shape resonances, the physically observable fingerprints of a pole flow discriminate these classes by whether the shape resonance remains mostly stationary.
We have also studied the interaction between a Feshbach resonance and an antibound state.
Here we found the magnetic position of the resonant feature and the bound-state pole to converge at threshold, replicating qualitatively the prediction of Ref.~\cite{Marcelis2004}. The bound state does not connect with the above-threshold resonance pole that we infer from a Pad\'{e} approximant, unlike the single-channel case for a weakened $\ell >0$ potential, where a bound state transitions continuously into a (shape) resonance \cite{Taylor2006}. This is in accordance with Ref.~\cite{Bortolotti2008} which for our particular system notes the resonance to appear ``well before the disappearance of the bound state''.

Resonance-pole interactions and BICs have been studied and engineered extensively in classical coupled-resonators and their utilisation is for example at the cutting edge in field of photonics~\cite{Kodigala2017,Kang2023}. In contrast, observations of BICs in the quantum scattering domain where it was first formulated~\cite{Neumann1929} have remained partially elusive. Here, we have predicted the emergence of a BIC in physically-realisable collisions of \Rb{87}---one of the species we have routinely studied with our laser-based collider. The identified BIC, however, appears at a comparatively high energy, where broadening that is integral to our acceleration scheme may preclude a straight-forward experimental observation and initial attempts at locating the lower branch of \figref{fig:bic_poles} have been unsuccessful. The detection scheme brought forward in Ref.~\cite{Horvath2017} might provide a route forward, but locating an alternative BIC candidate at lower energy would clearly be desirable.

}

\begin{acknowledgements}
This work was supported by the Marsden Fund of New Zealand (Contract No. UOO1923). NK acknowledges the hospitality of Aarhus University during the write-up of the manuscript.
\end{acknowledgements}


\end{document}